# Connectedness, Cognitive Load, and Human-AI Oversight in Cyber Operations

Platform Environment and Cognitive Load as a Cognitive Vulnerability

Nathan Conklin
Department of Computer Science
Virginia Tech
Blacksburg, VA, USA
nathan.conklin@vt.edu

Peng Gao
Department of Computer Science
Virginia Tech
Blacksburg, VA, USA
penggao@vt.edu

Chris North
Department of Computer Science
Virginia Tech
Blacksburg, VA, USA
north@cs.vt.edu

## ABSTRACT

AI-assisted cyber situational awareness triggers machine-generated reasoning traces (step-by-step justifications for anomaly classifications) that a human operator is expected to review. Because cyber signals and their traces arrive faster than any operator can process, human review is the limiting constraint on oversight. The standard approach is to identify the riskiest cyber events for review using model-side signals such as confidence or uncertainty. That framing ignores the operator's cognitive capacity which varies sharply with the operational environment. We propose an alternative where the system's environmental and connectivity telemetry serves as an available, non-invasive proxy for operator load. That same telemetry determines whether the human-AI partnership can reach the broader collective for support. In a maritime platform, environmental and connectivity attributes including depth, number of active communications paths, density of the tracked contact picture, and operational tempo all carry this signal. Need for operator oversight becomes a decision that materializes as a combination of both risk and environment-derived operator capacity. We present a reference architecture for a connectedness-aware oversight engine, demonstrating everyday use cases alongside its intended incorporation into the submarine cyber-defense toolkit. Two themes emerge: 1) the operator's environmental state is itself a connectedness measurement, and 2) connectedness drives the cognitive load and defines a collective boundary in human-AI cyber operations.

## 1 Introduction

Security operations are shifting from human-led triage toward human-AI teams in which machine agents investigate anomalies and emit reasoning traces for analyst review. Traces arrive faster than any human can read them, so the operator's attention is the binding constraint on oversight. The dominant response is risk-based routing: escalate high-risk cases, automate the rest, and decide between the two using model-side confidence and uncertainty. That approach works, but it assumes the reviewer brings constant capacity to every flagged case.

The assumption fails in the field, and its failure is a cognitive-security problem. Cognitive security concerns the integrity of the conditions under which people reason for themselves (El et al. 2026). An operator whose capacity has collapsed under environmental load cannot evaluate the data, its accuracy, or its usefulness; no attention remains to spend on the question. External stressors and the operational environment drive cognitive load, and load in turn degrades all five cognitive-security dimensions: attention, belief formation, memory, judgment, and agency.

In a submarine, the watchstander most exposed to this is the Fire Control Technician (FT), who operates and maintains the combat control systems and cybersecurity display for a platform of nearly a thousand nodes. Alerts flood the operator, many of them software misconfigurations: the sailor can fix some, others require depot action, and others ship with the system as installed and never clear. Alert fatigue follows, and the operator stops believing the display and stops attending to it. In the maritime case, the environment that raises the operator's cognitive load the most is also the environment in which a real intrusion does the most damage.

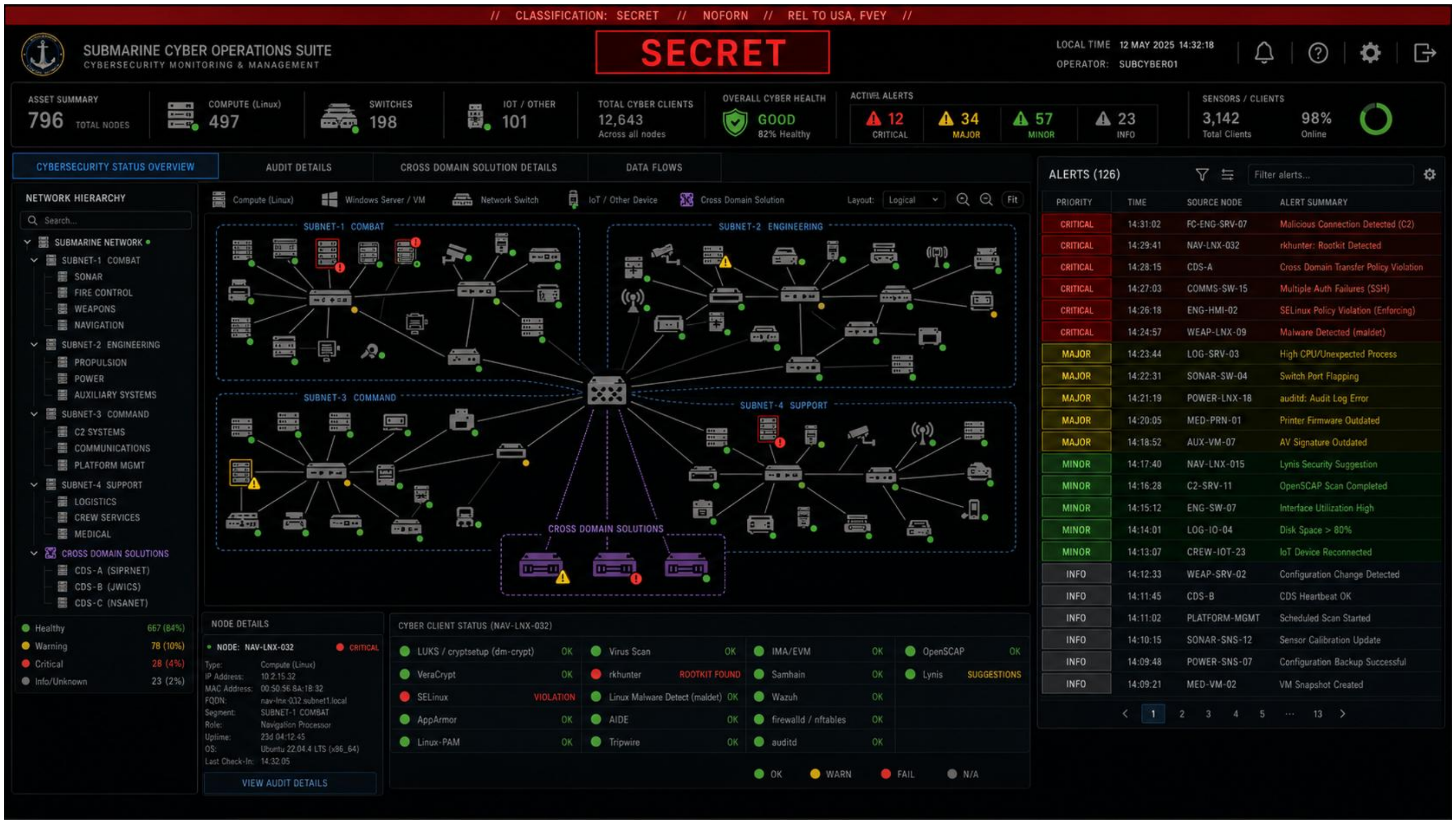

**Figure 1: Cybersecurity Display depicts the network and system status, the security threat posture, and active alerts that warrant operator review and action. (Generated by prompt shown in Appendix A). Figure is UNCLASSIFIED//DIST A.**

We make oversight a joint function of model-side risk and an environment-derived estimate of operator capacity, computed from telemetry the platform already produces. We (1) frame cognitive load as a cognitive vulnerability that degrades all five cognitive-security dimensions; (2) introduce the platform as a non-invasive sensor of its own operator; (3) specify the operational telemetry used to estimate load and distinguish it from cyber exposure; (4) present a reference architecture, the connectedness-aware oversight engine; (5) ground it in a Fire Control Technician use case; and (6) develop the dual signal roles of connectedness and cognitive load as an exposure signal and operational filter.

## 2 Background & Related Work

Cognitive load theory holds that working memory is narrowly bounded, on the order of four chunks under conditions that prevent rehearsal, and that exceeding this bound collapses processing rather than slowing it (Cowan 2001; Miller 1956; Sweller 1988; Sweller et al. 1998). External stressors compound the limit. Acute stress narrows controlled attention through attentional tunneling and degrades working memory, judgment, and cognitive flexibility; a moderate band of stress aids performance while an overwhelming band erodes it (Girotti et al. 2018; Klein & Boals 2001; Montañez et al. 2020). Alert fatigue is the operational signature of that collapse in alerting systems. Security operations center teams receive an average of 4,484 alerts per day, cannot action 67% of them, and report that 83% are false positives (Vectra AI 2023); analysts rationally learn to ignore alert classes that benign local behavior explains (Alahmadi et al. 2022; Tariq et al. 2025). The pattern holds outside security. Invalid alarm rates in clinical settings run from 85% to 99.4% of all alarms (Burdick et al. 2023), ICU nurses face as many as 350 alarms per bed per day (Cvach 2012), and the resulting "cry-wolf" effect slows response to true events measurably (Bonafide et al. 2015; Drew et al. 2014). Those hazards led ECRI to name alarm hazards the number-one health-technology hazard (ECRI Institute 2011).

The trust-in-automation literature names two failure modes. Distrust and disuse follow when a system cries wolf; complacency and overtrust follow when reliable automation lulls a multitasking operator, an effect that intensifies under load and when reliability holds constant rather than varying (Lee & See 2004; Parasuraman & Riley 1997; Parasuraman & Manzey 2010). Adaptive automation reallocates work to hold the operator within a workable band of demand (Endsley & Kaber 1999; Kaber & Endsley 2004; Parasuraman et al. 2000), but that work has triggered reallocation from the operator's body, using EEG or heart rate (Aricò et al. 2016), or from task-error counts. Both routes demand instrumentation that is invasive, fragile, or operationally unacceptable on a submarine watch floor. Physiological load estimation reaches better than 90% accuracy in the laboratory using multimodal indices, but that accuracy depends on body-worn and eye-tracking sensors (Ma et al. 2024). A separate tradition estimates load from environmental context, such as traffic and road demand for drivers, without touching the operator (Meiser et al. 2022; Piechulla et al. 2003). Chain-of-thought monitoring offers a promising but fragile route to oversight; reasoning traces are frequently unfaithful to the computation that produced them (Korbak et al. 2025; Lanham et al. 2023; Turpin et al. 2023). The one longitudinal study of analyst-LLM collaboration in a live security operations center finds that analysts use models for sensemaking and context building rather than for final determinations, and it recommends surfacing evidence rather than recommendations (Singh et al. 2025). Our setting adds a constraint that extends this work: for mobile environments connectivity is intermittent by physics and by doctrine. In addition, communications connectedness determines cyber exposure and whether the operator can reach help.

## 3 Cognitive Load as a Cognitive Vulnerability

We treat cognitive load as a cognitive vulnerability in the catalog sense (Gupta et al. 2026). An AI affordance, a high-volume stream of machine-generated alerts and reasoning traces, meets a human limitation, bounded working memory under environmental stress, and the pair produces a failure pattern: alert fatigue and learned disregard of the display. This vulnerability corrodes oversight because it degrades all five cognitive-security dimensions at once.

***Attention***. High load narrows the operator's effective field and produces attentional tunneling, so the operator never perceives peripheral but genuine alerts; perception, the first level of situation awareness, fails with no outward sign (Endsley 1995). ***Belief formation***. Under a cry-wolf regime the operator forms and hardens the belief that the display is noise, and stops updating on its outputs. That inference is rational given a false-positive-dominated history, and catastrophic when the rare true positive arrives. ***Memory***. Load consumes the working-memory capacity the operator needs to hold the tactical picture and the network picture together, so cross-cutting patterns, such as a slow sequence of medium-severity alerts across subnets, never come together. ***Judgment***. Stress degrades the deliberate evaluation that separates a true positive from a benign misconfiguration, the discrimination that justifies keeping a human in the loop. ***Agency***. Chronic overload pushes the operator toward one of two surrenders of reasoning: disuse, in which the operator abandons the display, or automation deference, in which the operator rubber-stamps its outputs. Either path ends with the human no longer reasoning for themselves, the loss cognitive security exists to prevent (Frischmann & Selinger 2018; Parasuraman & Riley 1997; Risko & Gilbert 2016).

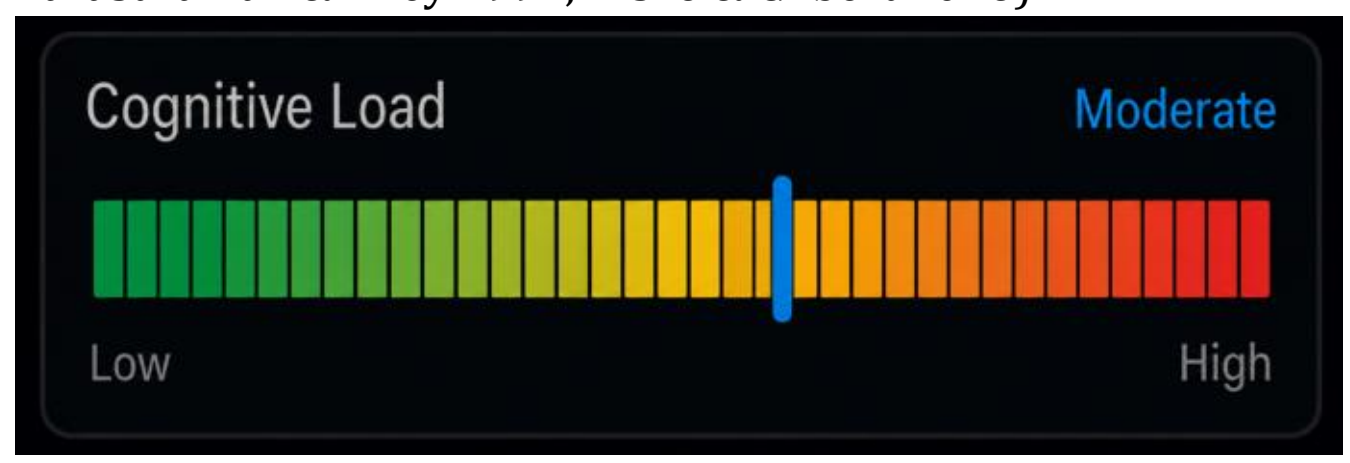

**Figure 2: The operator's Cognitive Load indicator. Cognitive Load is estimated non-invasively from platform environmental and connectivity telemetry.**

## 4 Fire Control Technician Use Case

Consider a Fire Control Technician standing watch on a submarine. The FT operates and maintains the boat's Combat Control System and the Submarine Warfare Federated Tactical System (U.S. Department of the Navy 2025), performing contact management, weapon planning, and synthesis of SONAR and imaging data. In the concept we develop, the FT also monitors the single aggregating cybersecurity display for the platform, a network of nearly a thousand nodes spanning combat, engineering, command, and support subnets (Fig. 1). The display presents asset health, a subnet hierarchy, a topology graph, and an alert queue, along with per-node findings such as a rootkit indicator or a mandatory-integrity-architecture violation. Those findings are the machine-generated reasoning artifacts the watchstander is nominally expected to review.

The alerts are heterogeneous in a way that defeats naive triage (Fig. 3). A malicious command-and-control connection and a detected rootkit sit in the same list as an outdated printer firmware notice, an expired antivirus signature, and a disk-space warning. The sailor can fix some alerts; others require depot-level action the crew cannot take at sea; and others are artifacts of the system as installed and will never clear. The FT responds as SOC analysts and ICU nurses do, learning which alert types are almost always benign and then closing or ignoring them. After that, the display intended to provide cyber oversight provides none; the watchstander reads it as background noise.

Cognitive load carries cyber-defense consequences on this watch. The FT's load rises most sharply in five states: self-protect events, periscope-depth or surfaced transit, hull-opening evolutions, a dense contact picture, and high maneuver tempo. Those are the same states in which additional reasoning traces compete with mission-critical work. The conditions that raise the FT's load, operating shallow, under emission control, through a congested passage with a dense contact picture and elevated tempo, are the same conditions under which a real intrusion does the most damage and is hardest to recover from, because the boat cannot surface, phone shore, and hand the problem to a reach-back cell (Burrell 2023; Meadors 2022). A model-side-only router, blind to the operator's state, would generate the same volume of notifications as it would during a quiet transit, delivering the appearance of oversight without its substance.

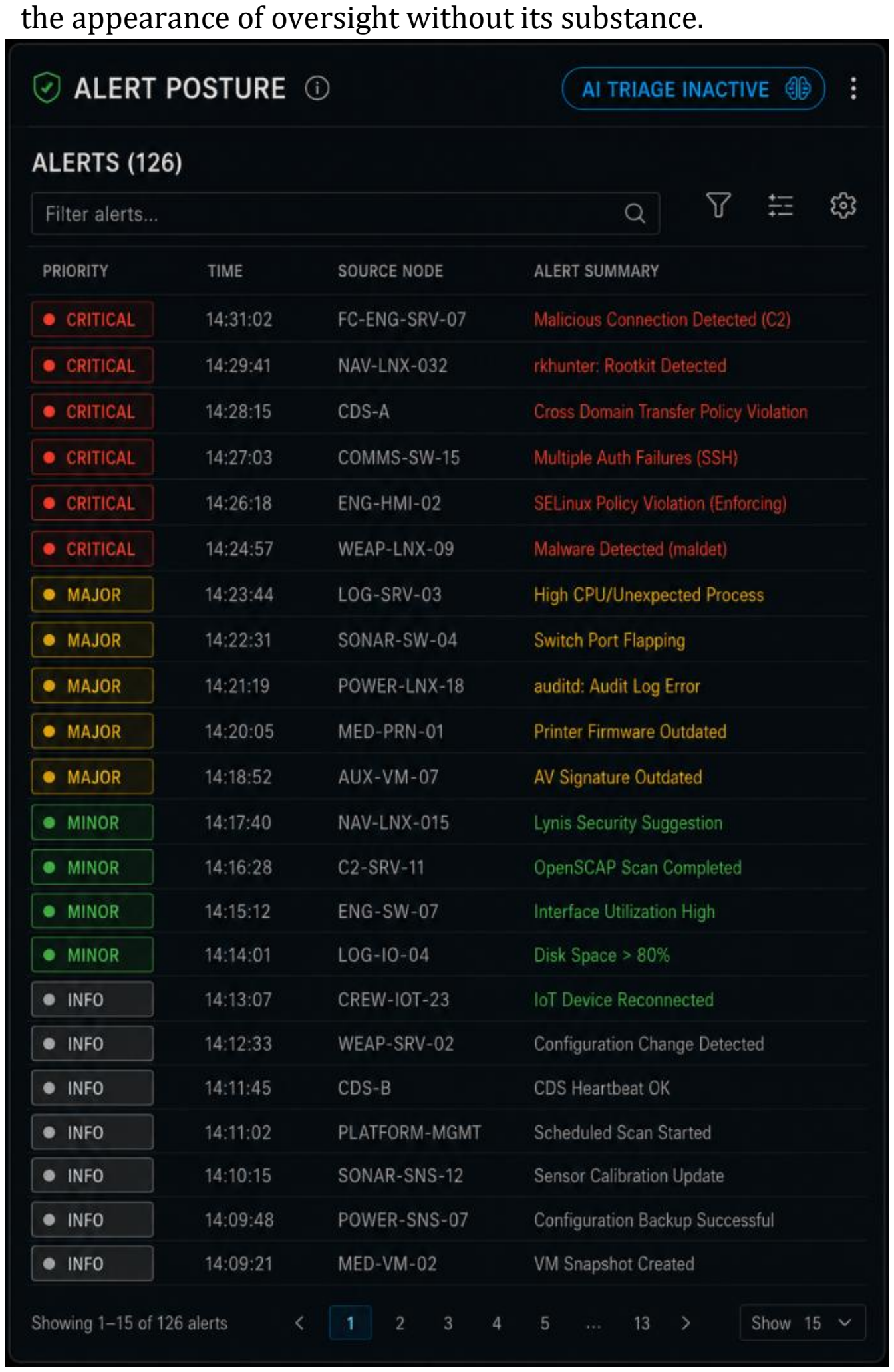

**Figure 3: The operator is overloaded with alert fatigue and begins to ignore the notifications.**

## 5 Telemetry-Derived Cognitive Load and Cyber Exposure

The combat system already records enough telemetry to estimate two separate operational quantities: room-level cognitive load $L(t)$ and cyber exposure $E(t)$. A rule-based estimator maps $L(t)$ inversely to the operator-capacity budget $\hat{C}(t)$; $E(t)$ represents the likelihood that current external-interface activity could produce a genuine cyber event. The policy keeps these quantities separate because they can move in opposite directions. High load supports deferring low-value notifications; high exposure supports lowering the escalation threshold. The estimator requires no wearable devices, biometric sensors, new emissions, or crew interaction. These rules are candidate operational heuristics and require calibration against recorded telemetry and human review performance.

The allocator evaluates $L(t)$ and $E(t)$ separately. A routine communications window creates low load with high exposure, so the policy should lower the escalation threshold. Torpedo evasion without external connectivity creates high load with low exposure, so the policy should raise the threshold for low-value alerts. When both are high, alert importance and the availability of connective boundary $B(t)$ determine whether the system suppresses, interrupts, or escalates. Future evaluation should test casualty states, close-aboard or collision-avoidance conditions, time-critical weapon and countermeasure countdowns, and other telemetry that reflects sustained crew commitment.

**Table 1: Candidate telemetry rules for estimating cognitive load and cyber exposure.**

| Signal class | Telemetry rule | Interpretation / policy use |
|---|---|---|
| **Safety / self-protect** | Weapon powered; man overboard; torpedo in water; countermeasures display active. | Any listed state sets $L(t)$ to High; emergency or self-defense tasks consume the crew's attention. |
| **Depth regime** | Periscope depth or surfaced transit: High. Surfaced communications or dockside: Medium. Deep: Low. | The estimator assigns load from the operational demands of each depth regime. |
| **Contact density** | >15: High. 5-15: Medium. <5, or all contacts beyond approximately 3 nmi: Low or normal. | Contact count and proximity estimate the crew's deconfliction burden. The overlapping thresholds require calibration. |
| **Hull openings** | Torpedo tube or dry deck shelter open: High. Surfaced sail hatch: no signal. | Torpedo-tube and dry-deck-shelter openings indicate weapons or special-operations activity; routine surface access does not. |
| **Maneuver tempo** | Three or more own-ship maneuvers during the preceding hour: High. | Maneuver rate, rather than a single maneuver, indicates that the crew is working a tactical problem. |
| **Training activity** | Training displays or simulated data active: Low. | Training activity anchors the low end of $L(t)$. |
| **Cyber exposure $E(t)$** | Active communications or external-interface activity: elevated $E(t)$, not $L(t)$. | $E(t)$ lowers the escalation threshold based on $B(t)$; it does not change $L(t)$. |
| **Excluded signals** | Routine displays; navigation; mission products or phase; manual overrides; display-change rate; non-cyber alert volume. | Most are non-discriminating; display-change rate and alert volume may be adversary-influenced or attack-correlated. |

## 6 Connectedness-Aware Oversight

The connectedness-aware oversight engine wraps an existing AI cyber-analysis capability and adds three components: a per-trace risk estimate $R(\tau)$, an environment-derived capacity estimate $\hat{C}(t)$, and an allocator that joins them. $R(\tau)$ draws on the established routing literature: model confidence and uncertainty, faithfulness proxies, and chain-of-thought structural features such as trace length. Trace length predicts both task difficulty and human review time, so it also feeds the capacity budget as a review-cost estimate (Davidson 2025).

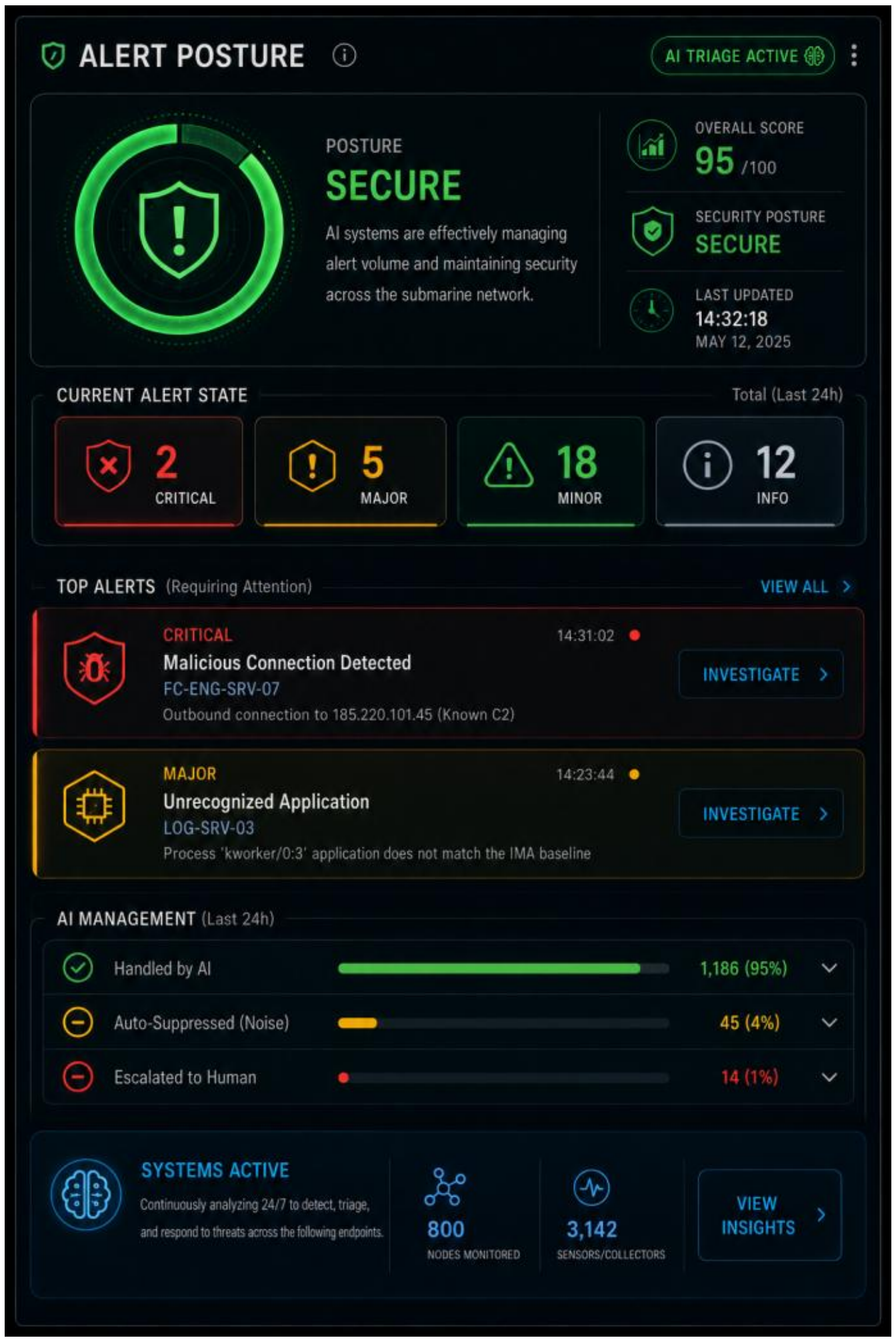

**Figure 4: AI-Assisted alert management improves cyber situational awareness by utilizing the operator cognitive load to control the notification management.**

$\hat{C}(t)$ is the novel pathway. The engine computes it from platform telemetry: depth band; count and status of active communications paths; tracked-contact density; weapons- and training-display state; operational tempo; and the alert volume itself. It renders the result to the watch as the Operator Cognitive Load indicator (Fig. 2). This is the "platform as a sensor of its operator" claim, and the sensing stays non-invasive by design: no electrodes, no eye trackers, nothing that requires consent, hygiene, or recalibration at a watch change. The connectivity components of the telemetry do double duty by also defining a collective boundary $B(t)$. With communications available, shallow with mast or buoy deployed and emissions permitted, the engine can escalate an ambiguous high-risk trace off-hull to a shore cyber cell, and the dyad operates as a node in a larger collective. Deep and emissions-constrained, the dyad cannot reach that collective and must self-resolve. The same physical facts that raise the operator's load also close the boundary, which makes connectedness a single variable serving as both the load signal and the collective boundary.

The allocator selects a review action per trace from a graded set: auto-handle and log (d0); spot-check the single highest-risk step (d1); full audit of the complete chain (d2); or escalate to the collective (d3), available only when $B(t)$ permits. It spends the capacity budget $\hat{C}(t)$ on the most valuable review actions first, raises the bar for what reaches the human as capacity falls, and drops d3 from the action set when the dyad cannot reach the collective. At high capacity the operator reviews more traces directly, which preserves skill and calibration. At low capacity the AI absorbs more of the reasoning, and only the highest-risk traces reach the operator, presented as highlighted single steps rather than full chains. The AI Management view (Fig. 4) makes the division legible: the routine majority handled by the AI, a slice auto-suppressed as noise, and a small set escalated to the human, with the auto-handle threshold moving as measured load rises and falls.

Raising the threshold under high load appears to quiet the display in the states that Section 4 identifies as highest consequence. Load and exposure move independently by design: high $L(t)$ raises the threshold only for low-value traces and elevated $E(t)$ lowers it regardless of load.

## 7 Discussion

The concept engages both of cognitive security's reliance failures. Under-reliance, or disuse, is the status quo the engine is designed to repair: the fatigued FT who ignores a display that cries wolf. Over-reliance, or complacency, is the risk the engine itself introduces, because raising the auto-handle rate under load can train the operator to rubber-stamp, and complacency intensifies under the same multitask load we are addressing (Parasuraman & Manzey 2010). We therefore treat two features as requirements. First, the engine retains every auto-handled trace for later audit, so a reviewer can inspect the decision to hide a trace from the human. Second, periodic forced-review sampling reinserts human review by design; that friction keeps the operator calibrated and holds the automation to a variable rather than constant reliability profile. Appropriate reliance, more than raw automation accuracy, mediates team performance, and the largest gains appear in the complex, uncertain tasks where human and machine error patterns diverge (Gonzalez et al. 2026; Schemmer et al. 2023).

The contribution maps onto the CVMC schema as a catalog entry (Gupta et al. 2026). **Affordance**: high-volume AI-generated alerts and reasoning traces under intermittent connectivity. **Cognitive vulnerability**: environment-driven operator cognitive load. **Failure pattern**: alert fatigue, cry-wolf disuse, and complacent rubber-stamping. **Mitigation**: connectedness-aware allocation of review depth with load-adaptive presentation. **Evaluation metric**: catch-rate-per-unit-capacity, miss rate, time-to-disposition, and calibrated reliance (aligned with the workshop's over-/under-reliance measurement agenda). **Burden-and-equity check**: the mechanism must not relocate risk into an unauditable gate, and its load model must not systematically disadvantage watchstanders whose baseline capacity differs. **Governance hook**: the audit log over hidden traces, together with the forced-review sampling policy, provide the inspection surface an oversight mechanism of this leverage requires.

The core empirical claim, that environmental telemetry predicts measurable degradation in human review performance, is a falsifiable hypothesis rather than a result. If environment adds nothing over model confidence, the contribution narrows to the routing baseline; we would treat a clean null as publishable given the prevailing assumption that human capacity is constant. The framework generalizes to settings that instrument the environment more readily than the operator, where the environment drives load and, for the collective-boundary half of the argument, where connectivity is intermittent. Satellite and spacecraft operations, disaster-response posts, and industrial control all qualify. The submarine sits at the strong end of that spectrum because both conditions hold at once and by physics.

## 8 Conclusion

Oversight of AI reasoning is only as good as the human's capacity to perform it, and in operational settings the environment produces that capacity and the platform already measures it. Treating the platform as a sensor of its operator, and connectedness as both a load signal and a collective boundary, lets us route scarce human attention where it protects the mission; it also removes the false comfort of routing that assumes a tireless reviewer. Framed as cognitive security, the goal is to preserve the conditions under which the watchstander can still reason, believe, remember, judge, and act for themselves.

## ACKNOWLEDGMENTS

Thanks to Greg Caufman and Mason Baron for your support and enabling this research. Thanks to Meaghan O'Rourke, Chris Crogan, and Nate Broadhurst for constantly challenging the ideas while supporting the best work.

## APPENDIX A

The following prompt was used to generate Figure 1. Example figures are shown as they are DIST A releasable. Actual system figures are either CUI or SECRET and not publishable in an open research forum.

---

Can you please create a picture of a software user interface that represents a system that is used by a submarine officer to manage all of the Cybersecurity alerts, cyber sensors/notifications, and cyber inspection activities for the ~800 nodes on the submarine system? The system includes nearly 500 compute machines (running Linux), nearly 200 switches, and roughly ~100 systems that are IoT devices (including Windows servers, VMs, printers, IP connected sensors, cross domain devices, etc.).

The display should be able to show all of the various nodes, each of them run roughly 3-24 cyber clients, and provide the status of the nodes and their cyber health. The system should also provide a layout of all of the alerts that are triggered throughout the system, the alerts are colored and prioritized (red is most important, yellow intermediate, green are least critical, and grey are notification/configuration items). The header at the top of the screen should have a Classification banner and should say "SECRET" within a red box (the typical coloring of SECRET classified systems). The alerts should be available on a panel on the right side of the screen to include a large list of alerts. There should be a portion of the screen w/ tabs to access the Cybersecurity status overview, audit details, cross domain solution details, and data flows. The main screen/UI that this picture is showing is the Cybersecurity Status overview tab.

The picture should have a dark theme, black background. Most of the status should be a lighter color (dark grey on a black background indicating the nodes are in good health, nothing to worry about). When nodes have individual alerts, there should be a visual indication that is colorized with the them (red/yellow/green) based on priority. Make the aspect ratio 16:9.

The cyber clients that provide status from the Linux machines include:
LUKS / cryptsetup (dm-crypt)
VeraCrypt
SELinux
AppArmor
Linux-PAM
Virus Scan
rkhunter
Linux Malware Detect (maldet)
AIDE — Advanced Intrusion Detection Environment
Tripwire
IMA/EVM
Samhain
Wazuh
firewalld / nftables
auditd
OpenSCAP
Lynis

---